\documentclass[preprint,12pt]{elsarticle}

\usepackage{amssymb}

\journal{Computer Speech and Language}

\begin{document}

\begin{frontmatter}



\title{Language Modelling for Speaker Diarization \\ in Telephonic Interviews}


\author[inst1]{Miquel India}
\author[inst1]{Javier Hernando}
\author[inst1]{Jos\'e A.R. Fonollosa}

\affiliation[inst1]{organization={Universitat Politecnica de Catalunya},
            addressline={C/ Jordi Girona 1-3}, 
            city={Barcelona},
            postcode={08034}, 
            country={Spain}}

\begin{abstract}
The aim of this paper is to investigate the benefit of combining both language and acoustic modelling for speaker diarization. Although conventional systems only use acoustic features, in some scenarios linguistic data contain high discriminative speaker information, even more reliable than the acoustic ones. In this study we analyze how an appropriate fusion of both kind of features is able to obtain good results in these cases. The proposed system is based on an iterative algorithm where a LSTM network is used as a speaker classifier. The network is fed with character-level word embeddings and a GMM based acoustic score created with the output labels from previous iterations. The presented algorithm has been evaluated in a Call-Center database, which is composed of telephone interview audios. The combination of acoustic features and linguistic content shows a 84.29\%  improvement  in terms of a word-level DER as compared to a HMM/VB baseline system. The results of this study confirms that linguistic content can be efficiently used for some speaker recognition tasks. 
\end{abstract}




\end{frontmatter}



\section{Introduction}

Speaker diarization addresses the problem of ``who spoke when'' in a multi-party conversation. Without prior knowledge of any speaker nor the number of the speakers in the speech, diarization aims to identify all the speech coming from each speaker. Two different sub-tasks can be distinguished in speaker diarization: speaker segmentation and speaker clustering. Speaker segmentation searches for the speaker turn boundaries, whereas speaker clustering aims to group all the speaker turns that correspond to each speaker. Depending on the speech domain, speaker clustering needs to determine the number of speakers in the audio. This work is focused on the telephonic domain, where it is assumed that only two speakers are talking.

The most common approaches used in speaker diarization are based on the Agglomerative Hierarchical Clustering (AHC) strategy. In this strategy, the system is initialized with a speech segmentation where each segment is assumed to correspond to one speaker. This initial segmentation can be created with different approaches like \cite{gupta2015speaker,bredin2017tristounet,yin2017speaker,jati2017speaker2vec, tranter2004speaker, tranter2006overview} or directly splitting the signal into homogeneous segments. AHC consists on  grouping iteratively these segments until each segment is assigned to its respective speaker. Therefore, in each iteration a pair of clusters is merged and a new segmentation is created in order to refine the speaker turn boundaries. The Bayesian Information Criterion (BIC) \cite{tranter2004speaker, tranter2006overview} was the conventional approach to decide which pair of clusters must be merged. Otherwise, Viterbi Decoding was the most used algorithm for the speaker re-segmentation. Speaker clusters were usually modelled either with Gaussian Mixture Models (GMM) or with i-vectors. I-vector framework \cite{dehak2011front} combined with  Probabilistic Linear Discriminant Analysis \cite{kenny2010bayesian, prince2012probabilistic} have shown a noticeable improvement in comparison with GMM approaches. This improvement has been shown for speaker clustering \cite{desplanques2015factor, sell2014speaker, dupuy2012vectors,shum2011exploiting, woubie2016improving} but still not for the segmentation task.

Deep learning has also been successfully applied for speaker diarization \cite{zajic2017speaker,le2017triplet, dimitriadis2017developing, wang2017speaker, sun2019speaker, huang2020speaker, wang2020speaker, flemotomos2019linguistically}, with different approaches in both clustering and segmentation tasks. Long Short-Term Memory Networks (LSTM) have been efficiently used to detect speaker turns boundaries either using acoustic features like in \cite{bredin2017tristounet,yin2017speaker, wisniewksi2017combining, yin2018neural, lin2019lstm}, or combining acoustic and linguistic content \cite{india2017lstm, park2018multimodal, park2019speaker, el2019joint}. On the other hand, the success of speaker embeddings for  speaker verification has led to use this approaches for clustering. 
This representation \cite{snyder2016deep,garcia2017speaker, diez2019bayesian, wan2017generalized,wang2017speaker} has been explored for the clustering task, outperforming i-vectors when a lot of speech data is available.

Natural language processing is one of the research fields where deep learning have caused a bigger impact. Neural networks have led to big improvements on analyzing and understanding natural language data. The most recent methods to extract features in tasks like machine translation, data mining or natural language modelling are based on word embedding approaches. Word embeddings are numerical word representations trained to capture the contextual information of a language \cite{mikolov2013distributed,mikolov2013efficient}.
Several models are known to produce these vectors, from the word2vec work presented in \cite{goldberg2014word2vec} to character-level models such in \cite{kim2015character}.  Word embeddings have shown its best performance in both language modelling and machine translation tasks when they are used as inputs of Recurrent Neural Networks (RNN)\cite{mikolov2010recurrent} or Transformers \cite{vaswani2017attention}. Works like \cite{india2017lstm, kim2015character, costa2016character, sundermeyer2012lstm, peters2018deep, howard2018universal, serban2015text}, exhibit the good performance of these embeddings with RNN architectures. Transformer based approaches like \cite{devlin2019bert, radford2019language, dai2019transformer, yang2019xlnet, brown2020language} have also shown state of the art results in NLP using these words representations.

In this paper we propose an alternative architecture  for speaker diarization in telephonic interviews, where our main contribution is a straight-forward algorithm that combines acoustic and linguistic information. The proposed approach is considered to be used for telephonic conversations, therefore the number of speakers per audio is known in advance. Additionally, our approach classifies each of the speaker clusters with an interviewer or customer label. Although there is a lot of tasks where linguistic content and speech have been successfully combined, the joint use of these sources has still not been fully explored for speaker diarization. Moreover, in several real-life applications it is needed to implement both Automatic Speech Recognition (ASR) and diarization \cite{canseco2004speaker, canseco2005comparative}, which increases the motivation on combining both systems.  Call-Centers have a wide set of tasks with different scenarios where is needed to perform call-transcription. This paper will be focused on the telephonic interview scenario which is a very important case for some Call-Centers. In this scenario, speaker patterns can be extracted from linguistic content in an easier way than in other cases, because part of the speech of some speakers may be prior known. In fact, the interviewer questions are commonly known and customers speech is sometimes limited to specific sets of expressions or answers such as giving a score, say yes or no, and so on.  Given this motivation, this work aims to research how to combine efficiently acoustic and linguistic data for speaker diarization in this scenario.  Therefore, in this work we present a LSTM based system where acoustic features are fused with linguistic content to identify the speech coming from different speakers. LSTM networks are commonly used in language modelling tasks to predict a word given the sequence of the previous words. In this work, we will use LSTMs similarly in order to infer about the speaker who says the word. With the possibility of adding acoustic features in the network, we examine its behaviour in a scenario where linguistic content contains discriminative speaker information. This scenario is based on a real application situation, more specifically in the Call-Center context.  Call-Center dialogues are composed by an operator-customer conversation where some part of the operator speech may be known a priori and the number of speakers is always known. 
In this work, our approach is evaluated on a database composed by telephone conversations where some interviewers make a survey to different customers. 
Given prior knowledge of the set of questions in each survey and the number of speakers in the conversation, the objective of this task is to identify the speech of the interviewer and the client interviewed for each recording.

The rest of this paper is structured as follows. Section 2 illustrates the architecture of the system. Section 3 gives the details of the system setup. Experimental results are presented in section 4. The concluding remarks and some future work are given in section 5.

\section{Architecture Description}
\label{sec:architecture}

The proposed algorithm is designed to work in a scenario where linguistic data contain speaker patterns. In this context, each recording is a two-speaker conversation where a first speaker (Interviewer) interviews a second speaker (Customer). These interviews are based on a survey composed by a set of questions which are similar for all the recordings. Therefore, the aim of the task is to find when the Interviewer and the Customer are speaking in each recording. The presented system uses both acoustic and linguistic content as inputs, hence the speech signal is initially pre-processed with an acoustic feature extractor and an ASR system. The output of the ASR and the acoustic descriptors are then used as inputs of the system. Given these inputs, the system will be trained to tag each word with its respective speaker label (Interviewer, Customer).

\begin{figure*}[!t]
\centering
\fbox{\includegraphics[width=\textwidth]{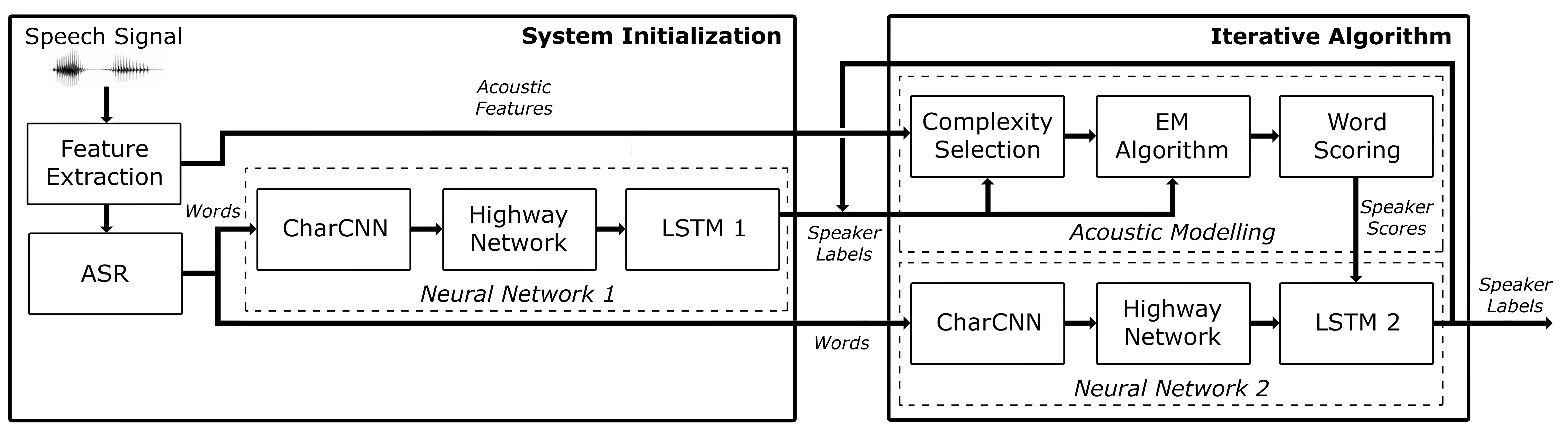}}
\caption{System Diagram.}
\label{Diagram}
\end{figure*}

The architecture of the proposed system is based on the iterative algorithm shown in Fig. \ref{Diagram}. Two different networks are used, each of them fed with different inputs. The system is initialized trough \textit{Neural Network 1}, which only uses linguistic content to create the first speaker labels for the iterative algorithm. On the other hand, \textit{Neural Network 2} works iteratively with both acoustic and linguistic data as inputs. Both networks work with sequences of word level representations and output the speaker labels corresponding to those input sequences of words. In each iteration the output speaker labels from the previous iterations are used to create two speaker models (Interviewer, Customer), which are used to extract an acoustic speaker score from each word. These scores indicate whether that word corresponds to the Interviewer or to the Customer. Hence, at each iteration of the algorithm, the speaker labels of the previous iteration are used to create the acoustic speaker scores which will be input additionally to the words in \textit{Neural Network 2}. The algorithm is iteratively run for a few iterations.

These neural network architectures are based on the system presented in \cite{kim2015character}. In our work, instead of using LSTMs to predict words, the networks are trained to tag each word with its corresponding speaker. The proposed algorithm proceeds by the following steps:

\begin{enumerate}

\item The system is initialized extracting both acoustic features and linguistic content. The words extracted from the ASR are introduced in  \textit{Neural Network 1}. These words are mapped into word embeddings (section \ref{sec:word}),  which are the input to the first LSTM. \textit{LSTM 1} yields an initial set of speaker labels (section \ref{sec:LSTM applied}) which will be used for the acoustic speaker modelling block in the iterative algorithm.

\item Given the speaker labels either from  \textit{Neural Network 1} in the first iteration or from \textit{Neural Network 2} in the next iterations, the two speaker acoustic models (GMMs) are created. These models are used to extract an acoustic speaker score for each word. These scores are calculated as the posterior probability of the Customer speaker GMM word given the word acoustic features (section \ref{sec:speaker}).

\item Acoustic speaker scores are used additionally to the words as inputs of \textit{Neural Network 2}. In \textit{Neural Network 2} the words are mapped into word embeddings and the concatenation of each word embedding and its acoustic speaker score is input to \textit{LSTM 2}. The output speaker labels  
from \textit{Neural Network 2} will be then used again in step 2 in a new iteration. The algorithm finishes after a few iterations and the last iteration output corresponds to the final result.

\end{enumerate}


\begin{figure}[!t]
\centering
\includegraphics[width=3.4in]{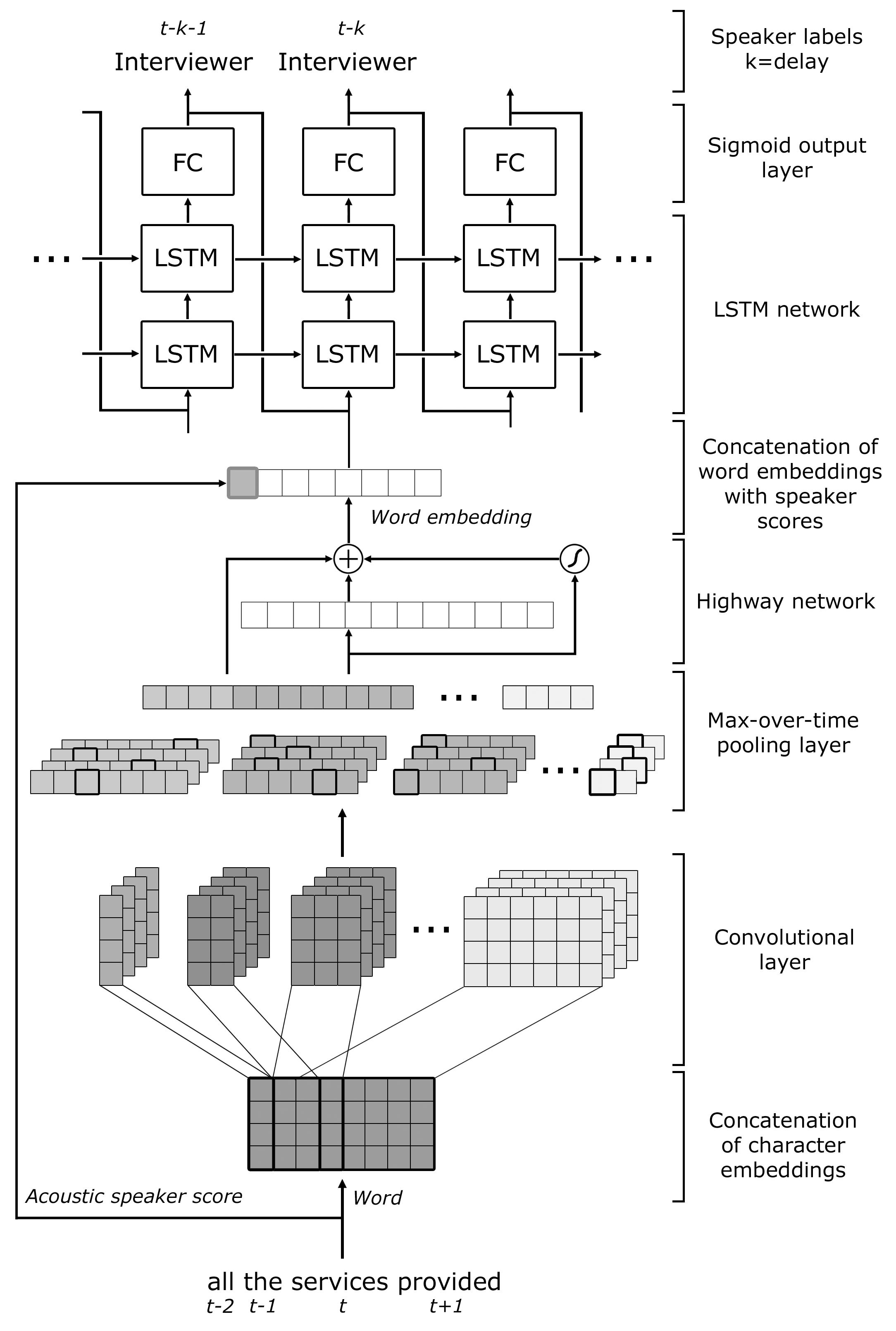}
\caption{Network Architecture Scheme. In Neural Network 1 the words are the only input. In Neural Network 2 acoustic speaker scores are input additionally in concatenation with word embeddings in the LSTM.}
\label{LSTM}
\end{figure}

\subsection{Character-level Word Embedding}
\label{sec:word}

The architecture of the proposed system (Fig. \ref{LSTM}) contains a LSTM neural network. This recurrent neural network uses as input sequences of word embeddings. Word embeddings are word representations modelled as real value vectors mapped from its textual form. In this system word embeddings are obtained from the output of a character-level convolutional neural network (CharCNN)\cite{kim2015character}.

In any language we can define a dictionary $V$ where each word can be represented as a vector $w$ in $V \in \mathrm{R}^{d'\times |\mathrm{C'}|}$. Variables $d'$ and $C'$ correspond to the vector and dictionary size, respectively. On the other hand, any word $w$ can be constructed as a sequence of characters $\left[ c_1,...,c_l \right]$, where $l$ is the word length. Therefore, if we define a dictionary of characters $Q \in \mathrm{R}^{d\times |\mathrm{C}|}$, where each character of a set $C$ is represented as a $d$ size vector, then any word can be constructed as a matrix  $\mathrm{C}^w$ $\in$ $\mathrm{R}^{d\times l}$. Character-based word embedding approaches map these 2D word representations into another low dimension space, which is discriminative in terms to the factor aimed to infer. Given a word $w$, a narrow convolution is applied between its representation $\mathrm{C}^w$ and a filter H $\in$ $\mathrm{R}^{d\times u}$ of width $u$. Applying a non linear function in the sum of this convolution and a bias term, we obtain a feature map $f^w$ $\in$ $\mathrm{R}^{l-u+1}$. The $i$-th element of $f^w$ is defined as:
\begin{eqnarray}
f^w[i] = \tanh(\langle C^w[*,i:i+u-1], H \rangle + b) 
\end{eqnarray}
where $C^w[*,i:i+u-1]$ is the $i$-to-$(i+u-1)$-th column of $C^w$ and $\langle A,B \rangle = Tr(AB^T)$ is the Frobenius inner product. We apply a max-over time pooling over the feature map so as to take the most representative feature in the vector: 

\begin{equation}
y^w=\max_i f^w[i] 
\end{equation}

where $y^w$ is the feature corresponding to the filter H  (when applied to the word $w$). Thus if we had a set of $N$ filters in the network, for each word $w$ we obtain a $N$ size representation $y = [y^{w1}, ..., y^{wN}]$, where each component is the output feature of a filter. For many NLP tasks the number of filters $N$ is used to be chosen between [100,1000]. 

Additionally to the CharCNN, one more network is implemented replacing $y_t$ with $x_t$ at each step in the LSTM architecture. Instead of using a typical set of fully-connected layers, those are replaced by a Highway network \cite{srivastava2015highway,srivastava2015training}. Highway networks are gate-based layers inspired by LSTMs, which have shown state of the art results in language modelling tasks \cite{kim2015character,jozefowicz2016exploring,zillyrecurrent}. Therefore, instead of using a feed-forward layer, $x_t$ is computed with the following function:

\begin{equation}
x_t =  T \odot g (W y_t + b) + (1-T) \odot y_t
\end{equation}

\begin{equation}
T =  \sigma (W_T y_t + b_T)
\end{equation}

where g is a non-linear function, $\odot$ is the element-wise multiplication, 
$T$ is the \textit{transform gate} and $1-T$ is the \textit{carry gate}. These layer gates allow to control whether each component of $x_t$ is obtained by a feed forward layer $g (W y_t + b)$ or it is directly carried from the input $y_t$.
As is shown in \cite{kim2015character}, these networks show better performance by modelling the interactions between the character n-grams extracted by the filters over $y_t$. Highway networks architecture was addressed to solve the learning issues found in large and deep networks. However, these networks are implemented in this system as  an alternative of deep feed-forward networks with the aim of optimizing the data flow across the layers.

\subsection{LSTM Word Classifier}
\label{sec:LSTM applied}

LSTM networks are used in this work in order to assign for each word its corresponding speaker. As is shown in Fig. \ref{LSTM},  the network assigns the speaker label to the corresponding introduced word, given its word representation $x_t$.  \textit{Neural Network 1} LSTM uses only as input word embeddings, meanwhile \textit{Neural Network 2} LSTM is fed with the concatenation of word embeddings and their respective acoustic scores. In our approach we use a two hidden layer LSTM network. Hence, the hidden state $h_t$ of the second LSTM layer is then used as input of a last dense layer, whose output corresponds to the speaker label $l_t$. In this case Customer and Interviewer label $l_t$ corresponds to outputs '1' and '0', respectively. Compared to the system presented in \cite{kim2015character}, we have implemented two extensions in the LSTM so as to adapt the network for this task:

\begin{enumerate}
\item Scheduled Sampling: In order to improve the model accuracy and the training stability, we have applied scheduled sampling \cite{bengio2015scheduled}. This method consists on using the previous output $\hat{l}_{t-1}$ as an additional input to $x_t$ in the LSTM during the training.  Hence in each training step, the LSTM input (Fig. \ref{LSTM}) will be the concatenation of $x_t$, $h_{t-1}$ and $\hat{l}_{t-1}$. Feeding the network with the groundtruth label leads to a faster convergence and a better model performance. In testing phase, $\hat{l}_{t-1}$ corresponds to the previous word speaker label. Therefore at time $t$ and considering a sequence of the past speaker labels $\left[l_1, ..., l_{t-1}\right]$ we extract both $t$ word Customer and Interviewer posterior probabilities. The inference is then posed as a decoding problem where we want to find the most likely sequence of speaker labels given the input sequence of words. 
We use the Beam Search algorithm to solve the speaker word decoding.
This approach is a Viterbi decoding variation which prunes the $K$ most likely hypothesis in each decoding step instead of considering all the paths.  

\item Output delay: Given a sequence of word embeddings $x = [x_1, ..., x_T]$, the inferred speaker label depends on the previous steps of the sequence but not on the next ones. Therefore, the network is trained with an output delay so the model decision in step $t$ depends not only on the past but also on $k$ future steps. In order to obtain the delayed desired label $l_t$, during training the network is then fed with the word embedding $x_{t+k}$, the hidden state $h_{t-1}$  and the desired output from the previous step $\hat{l}_{t-1}$.

\end{enumerate}

\subsection{Acoustic Modelling}
\label{sec:speaker}

Given the speaker labels either obtained from \textit{Neural Network 1} or \textit{Neural Network 2}, two acoustic speaker models are created (Interviewer, Customer) in each iteration. The MFCC features extracted in the system initialization are used to train a Gaussian Mixture Model (GMM) per speaker. We use the speaker labels to group all the features corresponding to the words of each speaker. These clusters are then used to train the GMMs using the Expectation-Maximization (EM) algorithm. The complexity selection of each speaker model is defined by means of the following expression: 

\begin{equation}
GM_j=\mathrm{round}\left( \frac{R_j}{CCR}\right)
\end{equation}

where the number of Gaussian mixtures $GM_j$ to model speaker $j$ is determined by the number of frames belonging to that cluster $R_j$ divided by the Cluster Complexity Ratio (CCR). CCR \cite{anguera2006automatic} is a constant value fixed across all the recordings that defines the number of frames needed per mixture in a GMM.

The two GMMs are then used to evaluate the set of words given each speaker model. For each word we extract a speaker score in order to refine the word labelling in each iteration. The score of each word is computed by the posterior probability of the Customer model given the features of this word. Hence, let define a word $w$ composed by a set of features $\left[ o_1,...,o_M \right]$, where $M$ is the number of frames in the word. The acoustic score is then computed as:

\begin{equation}
P(\mathrm{Cus}|w) = \frac{P(w|\mathrm{Cus})P(\mathrm{Cus})}{P(w|\mathrm{Cus})P(\mathrm{Cus})+ P(w|\mathrm{Int})P(\mathrm{Int})} 
\end{equation}

where Cus and Int refer to Customer and the Interviewer models and $P($Cus$)$ and $P($Int$)$ refer to their respective priors. Each speaker model $j$ is defined with a $\Omega_j$ GMM, composed by $GM_j$ Gaussian mixtures. The acoustic score of a word $w$ respect to the speaker $j$ modelled with $\Omega_j$ is defined as:

\begin{equation}
P(w|SPK_j)=\sum_{i}\log P(o_i|\Omega_j)
\end{equation}

\begin{equation}
P(o_i|\Omega_j)=\sum_{k}w_{jk}P(o_i|\Omega_{jk})
\end{equation}

where $P(o_i|\Omega_j)$ corresponds to the  $o_i$ ($i$-th frame assigned to $w$) likelihood given $\Omega_j$ GMM,  $P(o_i|\Omega_{jk})$ is the likelihood of 
$o_i$ given the $k$-th mixture of $\Omega_j$ and $w_{jk}$ is the corresponding mixture weight. The  posterior probability $P(Cus|w)$ of each word will be used as the acoustic speaker score input to \textit{Neural Network 2} LSTM.

\section{Experimental Setup}
\label{sec:setup}

The proposed system will be evaluated in a real Call-Center database against a conventional speaker diarization system. The details of the scoring metrics for this task, the database and baseline used and the system setup are given in this section.

\subsection{Scoring Metrics and Criterion}
\label{sec:scoring}

The most common metric used in speaker diarization is the Diarization Error Rate (DER). This metric considers three kind of different errors: Miss Speech (MISS), False Alarm (FA) and Speaker Error Rate (SER). The speech activity detection in this system is directly produced by the ASR system, where word-time stamps can be used as a very accurate speech segmentation. Hence the FA and MISS errors in our system are only produced by the ASR output and not by our diarization approach. For instance, in order to evaluate the performance of the presented approach, the FA and MISS are ignored and only the SER will be considered for the experiments. On the other hand, conventional DER is computed in terms of time duration. However, the algorithm presented works in word terms. Therefore, we have used a DER variation called Word-level Diarization Error Rate (WDER)\cite{park2018multimodal}. This metric is computed as the percentage of words that are assigned to a wrong speaker to the total number of words. 

\begin{figure}[!t]
\centering
\fbox{\includegraphics[width=0.9\textwidth]{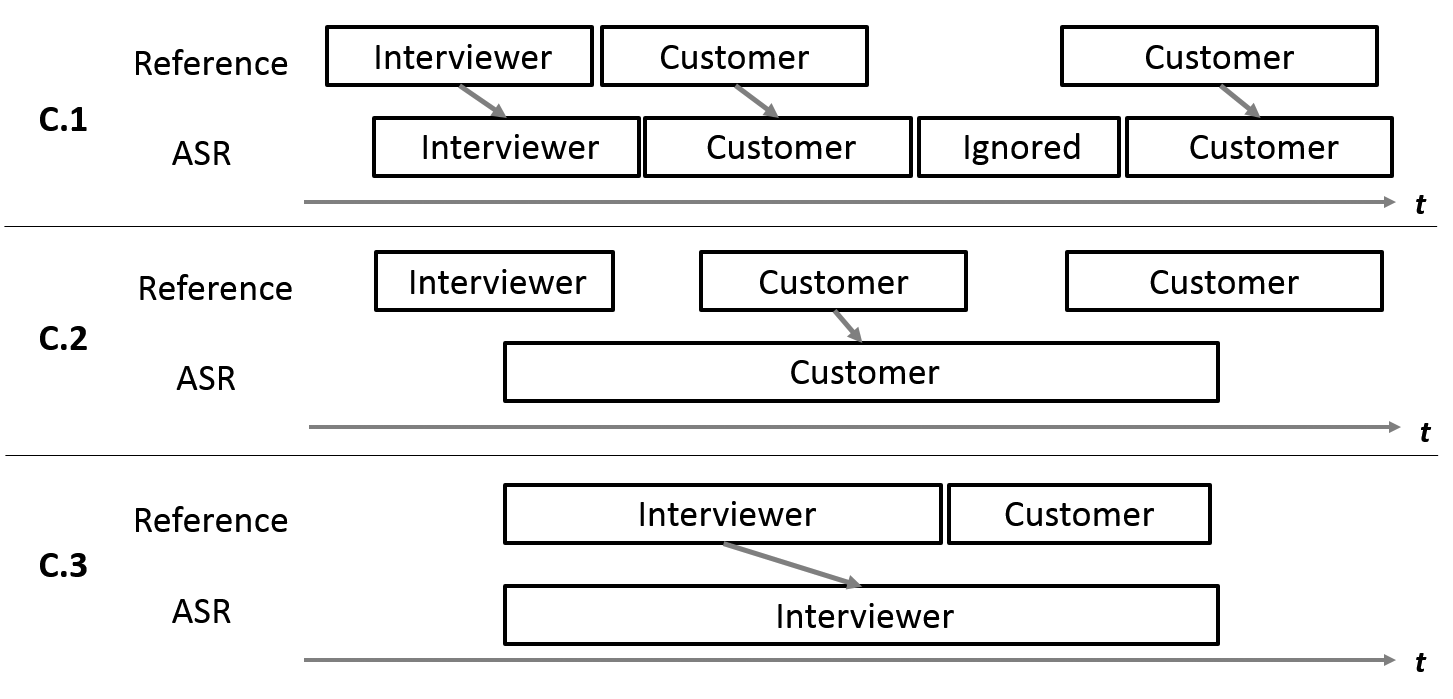}}
\caption{ASR Groundtruth Labelling: Boxes represent word segments with its respective speaker label. Arrows indicate the label assignation between transcription word labels to the ASR words. C.1, C.2 and C.3 correspond to condition 1, 2 and 3 in the direct overlapping criterion.}
\label{Conditions}
\end{figure}

In order to evaluate the presented approach, it is needed to use a modified reference that uses the same word segmentation as the one produced by the ASR. We have used a direct overlapping criterion so as to assign speaker labels from the manual transcription  to the word segmentation created by the ASR. Fig. \ref{Conditions} shows graphically how this criterion is applied according to the following conditions:

\begin{enumerate}
\item Given two time overlapped  transcription and ASR words, the transcription word  label is assigned to the ASR word if their overlap is bigger than half the time duration of the ASR word.

\item If 1) is not fulfilled but the overlap is bigger than half the time duration of the transcription word, the label is also assigned to the ASR word. 

\item If there is more than one transcription word overlapped to one ASR word. The label assigned corresponds to the word with the maximum overlap.

\item The ASR words that do not have transcription words time overlapped nor they fulfill the previous conditions are not evaluated. 
\end{enumerate}

\subsection{Database and Scenario Analysis}
\label{database:subsection}

\begin{figure*}[!t]
\centering
\fbox{\includegraphics[width=0.985\textwidth]{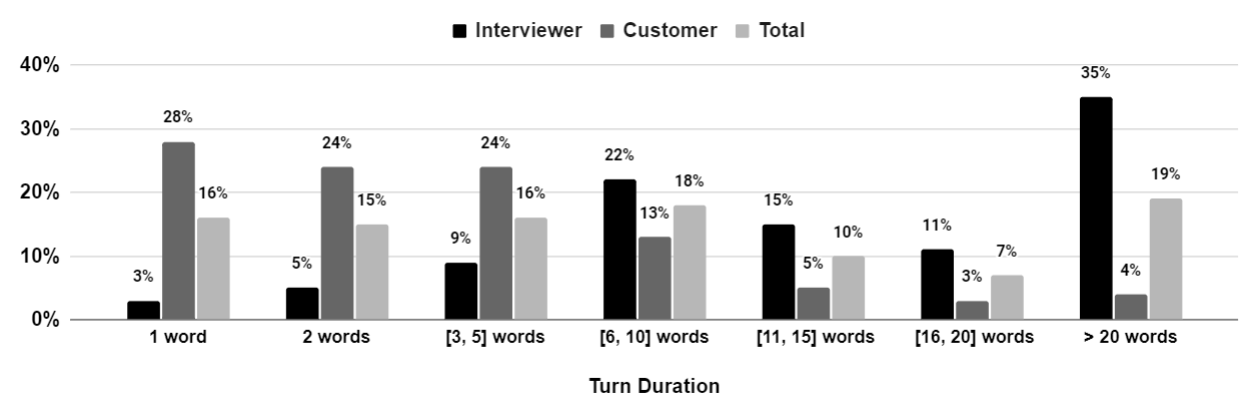}}
\caption{Turn duration distribution.}
\label{Histogram}
\end{figure*}

The database used for this work is a set of recordings from a Call-Center. This data has been obtained from a project with a private company, hence is not publicly available. Each of the recordings from this database contains a survey in Spanish of approximately 5 minutes duration. The survey context distinguishes two speakers: the Interviewer and the interviewed Customer. The questions asked in the survey are common for all the training and test recordings but with different speakers. This database is composed by 270 telephone recordings where  we used 240 for training and the other 30 for the test. This test partition is composed by a set of 18,299 words, where 14,498 words  correspond to the Interviewer and  3,801 words correspond to the interviewed people (Customer). The word labels are known from a manual annotation with its time stamps, where only word speaker labels are used for training. This manual annotation also contains special tokens which include noisy and overlap labels. These tokens have been removed for both training and test steps.

One of the main problems of this dataset is the unbalanced amount of speech signal between the two speakers. The interviewer speech comprise approximately more than the 79\% percent of the recordings. Furthermore, the interviewed Customer participation is reduced to short speech segments due to the content and the structure of the survey. Fig. \ref{Histogram} shows the turn duration distribution from the test partition. A considerable part of the interviewer speech is  based on the questions asked on the survey. Hence, their speaker turns are larger (more than 6 words) than the Customer ones, whose speech is mainly based on short answers (1 up to 5 words).  The lack of speech signal from the Customer prevents to perform a reliable diarization using only acoustic information.  In terms of clustering, speaker turns reduced to few words cannot be efficiently modelled with only acoustic features. In terms of speaker segmentation, cluster initialization is not accurately performed if is created splitting the signal into homogeneous segments. The unbalanced amount of speech of the two speakers produces that uniform initial segments are very likely to contain both speakers speech or only from the Interviewer one. Non uniform speaker segmentation approaches based on clustering methods did not also perform well due to the short speaker turns duration.

\subsection{Baseline}

With the aim of analyzing the impact of linguistic content in the proposed scenario for the speaker diarization task, we have selected a baseline that only models speaker traits from the audio features. The presented approach is compared with the Bayesian Hidden-Markov-Model (HMM) based algorithm proposed in \cite{diez2018speaker}. Given the database scenario and its conditions presented in the \ref{database:subsection} section, this algorithm has been considered as baseline due to its capacity to robustly estimate speaker models from very short speech segments. This system uses only acoustic features, works in the frame level and follows a Bayesian HMM topology. Each speaker state is represented as a low-dimensional vector $y_s$, given the following Joint Factor Analysis (JFA) based equation:

\begin{equation}
    \mu_s = \mu^{UBM} + Vy_s
\end{equation}

\vspace{5pt}
where given a Universal Background Model (UBM) trained as a GMM, the super-vector of concatenated Gaussian component means for a speaker s is posed as the sum of the UBM mean super-vector and the product of an eigenvoice matrix $V$ and its corresponding $y_s$ vector. This eigenvoice matrix $V$ is trained so as to project the speaker variability into a low-dimensional sub-space, although with this procedure the inter-channel variability is also modelled. Both UBM and $V$ matrix training and also $y_s$ extraction are described with more detail in \cite{dehak2011front, kenny2007joint}. Given this speaker modelling procedure, the Bayesian HMM topology is defined so as to assign for each speech frame its corresponding speaker state. In this HMM topology there can be multiple states per speaker, which all share the same specific distribution. Therefore a sequence of $D$ states can correspond to the same speaker imposing a minimum speaker turn duration. An iterative Variational Bayes (VB) based procedure is then used to infer about the speaker assigned to each speech frame. This algorithm allows to perform iteratively both segmentation and clustering tasks, where the stopping criterion is also defined through a VB based equation.

In order to evaluate the baseline with the word level scoring metric defined in \ref{sec:scoring}, we apply an overlapping criterion to tag each word given the frame labelling output from the baseline. Therefore, the label from the time overlapped frames to a word is directly assigned to that word. If there are overlapped frames from both labels, the label assigned to the word corresponds to the one with more overlapped frames.

\subsection{Optimization and Setup}

The ASR system used in our apporach to extract the words from the speech signal is based on \cite{povey2016purely}. We have implemented the ASR with the Kaldi toolkit \cite{Povey_ASRU2011} and its performance in the proposed database is about a 6\% Word Error Rate (WER), with a 1.4\% of insertions, a 0.4\% of deletions and a 4.2\% of substitutions. Following the same criterion applied to the manual annotation, special tokens have been removed from the ASR output. The neural networks were trained with the manual transcription words and tested with both transcription and ASR output words. On the other hand, the acoustic modelling was applied extracting MFCCs features. The extraction was done using 10ms shifted Hamming windows, where each frame contains 20 MFCCs coefficients. Hamming window length was set to 30ms. Speaker modelling was implemented by means of the EM algorithm, where the CCR ratio in order to define GMM mixtures was set to 7 seconds per Gaussian.

The two neural networks were trained by truncated back-propagation trough time \cite{werbos1990backpropagation,graves2013generating}. RMSprop was used with an initial 0.1 learning rate and the back-propagation was done for  35 steps. The learning rate was decayed by a 0.5 factor if validation perplexity did not improve by more than 1.0 after an epoch. Both networks were trained for 14 epochs with 20 size batches using binary cross entropy loss. For regularization we used dropout \cite{hinton2012improving} with probability 0.5. The dropout was applied on the LSTM input to hidden layers (except on the initial Highway to the LSTM layer) and the hidden-to-output sigmoid layer. Gradient updating was constrained to normalize gradient to 5. If the $L_2$ norm was above 5 in the batch, it was normalized again before the updating. 

Neural network architectures were setup similar to the large model presented in \cite{kim2015character}. The CharCNN was setup with a set of $h$ = 500 filters. These filters had the next range of widths $w$=[1,2,3,4,5,6], with the following number of filters per width [25,50,75,100,100,200] respectively. Character embeddings had a $d$=15 size and $tanh$ was the non-linear function applied in the convolutional step. The Highway network was set with only one hidden layer and Rectified Linear Units (ReLU) as activation functions. Both LSTMs were equally setup except for the speaker acoustic score introduced additionally in the \textit{Neural Network 2}. LSTMs were composed by 2 hidden layers, with 150 nodes per layer. Instead of using softmax-layer, the output layer was based on only one sigmoid activation with k=2 delay steps.

The baseline system was setup similar to \cite{diez2018speaker} but considering the proposed domain and the database size. We used 20 MFCC as features, we trained a 512 mixtures UBM-GMM with diagonal-covariance and the speaker latent variable $y_s$ size was set to 300. The VB inference setup is the same than \cite{diez2018speaker} except for D which was tuned to impose a 0.5 seconds minimum turn duration. Additionally, the system was tuned to directly force the algorithm to finish with two speakers.

\section{Results}
\label{sec:results}

The proposed approach has been thoroughly evaluated against the mentioned baseline in a Call-Center database. In order to analyze the individual and joint contribution of both acoustic and language modelling, two different outputs of the system have been evaluated. In one hand, the speaker labels from the \textit{Neural Network 1} output (NN1) will be used to analyze the performance of the system using only the linguistic content. On the other hand, the 
\textit{Neural Network 2} output will be used to evaluate the joint performance of both linguistic and acoustic features. We will consider the first iteration speaker labels (NN2) and the labelling when the system converges (NN2 (Iterative)). Furthermore, the WDER of both speakers (Interviewer, Customer) has also been computed in order to be more accurate in the results analysis.

Two different evaluations are presented in the following subsections so as to analyze the behaviour of the proposed approach. In section \ref{sec:main_results}, the different blocks of the system are evaluated and we analyze the performance of the algorithm when we use either the manual transcription words as inputs or the ones created by the ASR. In section \ref{sec:Speaker Turn}, the WDER of the systems will be evaluated for different speaker segment turn durations.

\subsection{Global Analysis}
\label{sec:main_results}

\begin{table}[!t]
\renewcommand{\arraystretch}{1.4}
\caption{WDER evaluation with different word input sources.}
\label{results}
\begin{center}
\begin{tabular}{c|c|c|c|}
\cline{2-4}
& \multicolumn{3}{|c|}{Oracle (Manual Transcription)} \\
\cline{2-4}
& Interviewer & Customer & Total\\
\cline{1-4}
\multicolumn{1}{|c|}{HMM/VB Baseline} & 3.22 & 50.04 & 13.05 \\
\multicolumn{1}{|c|}{NN1} & 2.99 & 10.08 & 4.47  \\
\multicolumn{1}{|c|}{NN2} & 1.68 & 4.34 & 2.23 \\
\multicolumn{1}{|c|}{NN2 (Iterative)} & \textbf{1.67} & \textbf{3.5} & \textbf{2.05} \\
\cline{1-4} \multicolumn{4}{c}{} \\ 
\cline{2-4} 
& \multicolumn{3}{|c|}{ASR} \\
\cline{2-4}
& Interviewer & Customer & Total\\
\cline{1-4}
\multicolumn{1}{|c|}{HMM/VB Baseline} & 3.5 & 51.32 & 13.55 \\
\multicolumn{1}{|c|}{NN1} & 3.51 & 13.44 & 5.34 \\
\multicolumn{1}{|c|}{NN2} & 1.74 & 5.04 & 2.35 \\
\multicolumn{1}{|c|}{NN2 (Iterative)} & \textbf{1.61} & \textbf{3.62} & \textbf{1.98} \\
\cline{1-4}
\end{tabular}
\end{center}
\end{table}

Table \ref{results} shows the WDER for the different systems with both input conditions: manual transcription (Oracle) and ASR. The baseline shows the worst performance in both conditions with a WDER higher than 10\%. On the other hand, the presented system shows a WDER lower than 6\% in both conditions for all the tested outputs. NN1 shows a 4.47\% and a 5.34\% WDER for Oracle and ASR conditions, respectively. Thus the proposed system outperforms the baseline with only using linguistic content as input.  NN2 has shown the best performance of all the evaluated systems. With only one iteration, using both acoustic features and linguistic content the system shows a 2.23\% WDER for Oracle conditions and a 2.35\% for the ASR ones. After a few iterations the best results are achieved with a 2.05\% and 1.98\% WDER for both Oracle and ASR conditions, respectively. Therefore, the combination of both acoustic and linguistic data provides the best results in the proposed scenario.

The training of both neural networks is done using the words from a manual transcription as inputs. However, in the testing phase we have evaluated the system with both manual transcription and ASR words. This evaluation has been done in order to analyze how the the word error introduced by the ASR decreases the system performance. As it is was previously shown in Table \ref{results}, the baseline performance is worst than the initial speaker labels produced by NN1 for both conditions. NN1 WDER shows a relative error improvement of  65.74\% in comparison with the HMM/VB system in the Oracle condition. In the ASR condition, NN1 also outperforms the baseline but with less margin. The relative WDER improvement between NN1 and the HMM/VB baseline is 60.59\%.  Although there is a system performance decrease caused by the WER from the ASR, NN1 still outperforms the baseline system using only linguistic content. Otherwise, we have also analyzed how the decreased performance produced by the ASR is less significant when we use acoustic data in the system. Iterative NN2 WDER shows a relative error improvement of 82.91\% in comparison with the HMM/VB system in the Oracle condition. In the ASR condition, this relative error improvement is similar with a 82.65\% in comparison to the HMM/VB system. Therefore, despite the word error introduced by the ASR, the use of acoustic data in the iterative algorithm leads to almost an identical performance when using the manual transcription as input. 

\begin{figure}[!t]
\centering\fbox{\includegraphics[width=0.8\textwidth]{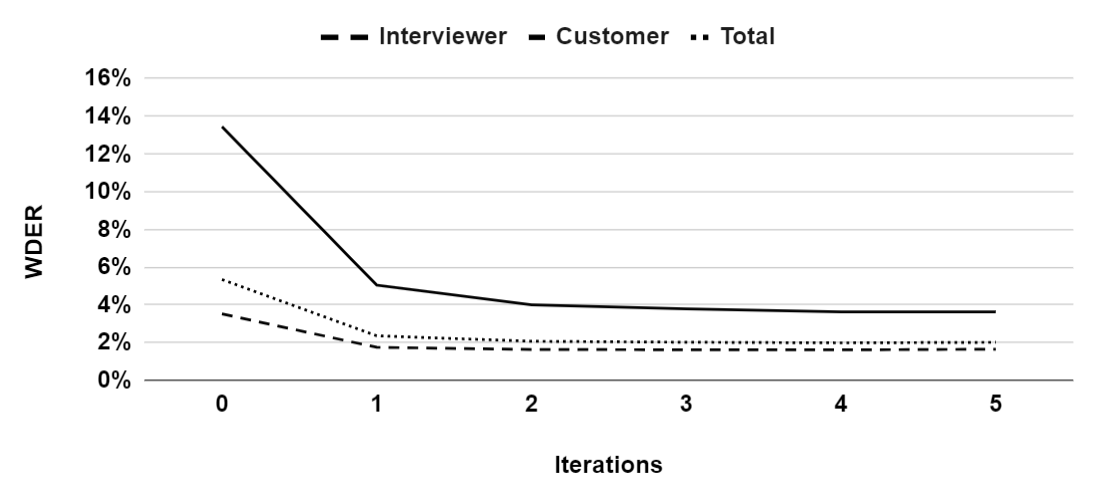}}
\caption{Iterative algorithm WDER parametrized by the number of iterations run in the system. The results shown correspond to the ASR condition. Iteration 0 corresponds to the initial speaker labels produced by \textit{Neural Network 1}.}
\label{Iterative}
\end{figure}

The iterative algorithm is initialized with the speaker labels provided by NN1. Hence, the number of iterations needed in the system to converge depends on the labelling produced by the system using only linguistic content. Fig. \ref{Iterative} shows the WDER of the iterative algorithm in relation to the number of iterations run for the ASR condition. In this figure we see that the system converges after 2 or 3 iterations. The speaker labelling created by NN1 corresponds to the iteration 0 with a Interviewer 3.51\% WDER and a Customer 13.44\% WDER. In the first iteration the WDER is decreased to 1.74\% and 5.04\% for both Interviewer and Customer, respectively. In the following iterations the system already converges around a 1.61\% Interviewer WDER and a 3.62\% Customer WDER. These results indicate that the initial speaker labels from NN1 are already very accurate. Therefore, NN2 only needs a few iterations to refine the speaker labelling with the addition of acoustic data. 

\subsection{Turn Duration Segment Analysis}
\label{sec:Speaker Turn}

\begin{figure*}[!t]
\centering
\fbox{\includegraphics[width=0.75\textwidth]{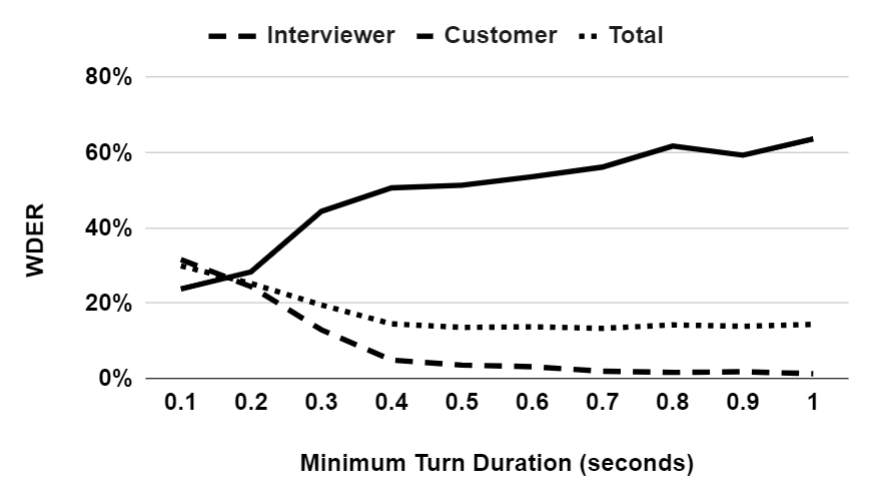}}
\caption{HMM/VB WDER parametrized by minimum turn duration applied on the model. The results shown correspond to the ASR condition.}
\label{Iterative_ASR}
\end{figure*} 

Conventional speaker diarization systems performance decreases when speaker turns are very short. The proposed baseline is based on a HMM topology that assumes a minimum turn duration in the model so as to avoid over-segmentation. This restriction increases the robustness of the system but also decreases the accuracy on the smaller segments.
Fig. \ref{Iterative_ASR} shows the WDER of the HMM/VB system in relation to the speaker turn duration condition applied on the model. As it is shown, there is a trade-off between the Interviewer and Customer WDER, which depends on the turn duration parameter. This trade-off is correlated to the average speaker turn duration of each speaker. In Fig. \ref{Histogram} is shown that most of the Interviewer segments have more than 6 words length and the Customer ones are shorter. Furthermore, there is more speech from the Interviewer than the Customer in the recordings. Therefore, if we decrease the minimum turn duration parameter, the Customer WDER increases but the Interviewer WDER decreases. This trade-off makes very difficult to set-up this kind of systems correctly.

\begin{figure*}[!t]
\centering
\fbox{\includegraphics[width=0.98\textwidth]{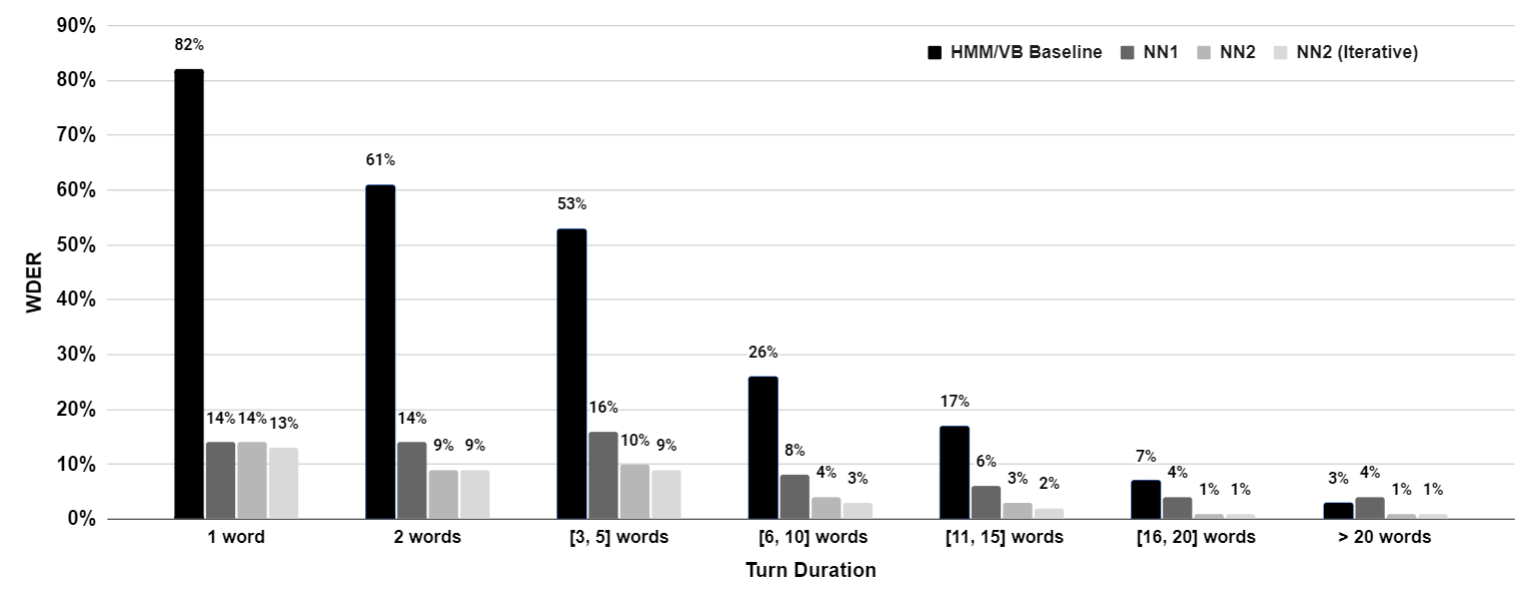}}
\caption{Total WDER evaluation for different speaker turn duration. These results are extracted for the ASR condition.}
\label{BIG_Graph}
\end{figure*}

Our proposed system infers directly over each word, hence is not needed to impose any temporary restriction. Table \ref{results} shows that the Customer WDER is still higher than the Interviewer one for all the experiments of the presented approach. However, the relation of both speaker errors is lower compared to the baseline system. In order to analyze the behaviour of the proposed system in different length turns, we have evaluated the WDER for different intervals of speaker segment durations. Fig. \ref{BIG_Graph} shows the total WDER of all the system blocks for several turn lengths.  As it was expected, the WDER increases as shorter are the turns for all the systems. The baseline system has more than 50\% WDER in turns shorter than 6 words. The proposed system outputs show better results in short turns with a global WDER between  9\% and 16\%. In turns larger than 6 words, the baseline system performance improves. Despite this improvement, our proposed system still outperforms the baseline for almost all the turn durations. Only the HMM/VB system shows better results in turns larger than 20 words compared to NN1, where only linguistic content is used. The benefit of using either only linguistic content or both linguistic and acoustic data in the system for different segment turns can also be analyzed from Fig. \ref{BIG_Graph}. The relative improvement between NN1 and Iterative NN2 for the total WDER is 7.14\% for 1 word turns and 35.7\% for 2 word turns. If we do the same analysis for long turns, the WDER relative improvement is about 75\% for both [16,20] word turns and  turns larger than 20 words. Thus acoustic data refines better the labelling produced by NN1 in the larger turns rather than in the shorter ones. Therefore, linguistic content can be efficiently used for tagging very short speaker turns, where acoustic data is less discriminative. On the other hand, the addition of acoustic data shows better results in larger speaker turns, where acoustic features are more effective.

\section{Conclusion}

In this paper we have investigated the combination of linguistic content and acoustic features for speaker diarization. We tested LSTM neural networks in order to merge acoustic and language modelling. This combination have outperformed the HMM/VB based baseline system where only acoustic data is used. Moreover, we have shown that language modelling is able to work better in situations where acoustic modelling performance is worse, such as in classifying short speech segments. The results indicate that with linguistic content, speaker diarization performance is less sensitive to decrease in short speaker turn conversations. For future work, it seems interesting to explore different acoustic based approaches that could perform efficiently with very short utterances. Additionally and considering that our work has only been tested for telephonic interviews, it would be also interesting to extend our approach to be used in scenarios with less correlation between linguistic content and speaker identities.

\section{Acknowledgments}

This work was partially supported by the Spanish Project DeepVoice (TEC2015-69266-P) and by the project PID2019-107579RB-I00 / AEI / 10.13039/501100011033.

\bibliographystyle{elsarticle-num} 
\bibliography{cas-refs}





\end{document}